\documentclass[conference, a4paper]{IEEEtran}
\IEEEoverridecommandlockouts
\usepackage{cite}
\usepackage{amsmath,amssymb,amsfonts}
\usepackage{algorithm}
\usepackage{algpseudocode}
\usepackage{graphicx}
\usepackage{textcomp}
\usepackage{xcolor}
\usepackage{url}
\usepackage{amssymb}
\usepackage{amsmath}
\usepackage{epsfig}
\usepackage[outdir=./]{epstopdf}
\usepackage{subfigure}
\usepackage{tabularx}
\usepackage{balance}
\usepackage{amsthm}
\usepackage{amsmath}
\usepackage{url}
\usepackage{amssymb}
\usepackage{amsmath}
\usepackage{epsfig}
\usepackage{epstopdf}
\usepackage{subfigure}
\usepackage{balance}
\usepackage{multirow}
\usepackage{booktabs}
\usepackage{color, colortbl}
\usepackage{tabularx}
\usepackage{soul}
\usepackage{authblk}
\usepackage{filecontents}
\usepackage[breaklinks]{hyperref}
\usepackage{url}
\usepackage{breakurl}

\DeclareMathOperator*{\argmax}{arg\,max}

\usepackage{color, colortbl}
\definecolor{LightCyan}{rgb}{0.88,1,1}

\usepackage{graphicx,subfigure}

\def\BibTeX{{\rm B\kern-.05em{\sc i\kern-.025em b}\kern-.08em
    T\kern-.1667em\lower.7ex\hbox{E}\kern-.125emX}}
\begin{document}

%
%
\title{Content-Aware Personalised Rate Adaptation for Adaptive Streaming via Deep Video Analysis}


\author[1]{Guanyu Gao\IEEEauthorrefmark{1}}
\author[1]{Linsen Dong\IEEEauthorrefmark{1}\thanks{\IEEEauthorrefmark{1}Equal contribution}}
\author[1]{Huaizheng Zhang}
\author[1]{Yonggang Wen}
\author[2]{Wenjun Zeng}
\affil[1]{Nanyang Technological University, Singapore}
\affil[2]{Microsoft Research Asia, Beijing, China}
\affil[1,2]{\{ggao001, lsdong, huaizhen001, ygwen\}@ntu.edu.sg, wezeng@microsoft.com}

\maketitle

\begin{abstract}
%
%
%
Adaptive bitrate (ABR) streaming is the de facto solution for 
achieving smooth viewing experiences under unstable network conditions.
%
%
However, most of the existing rate adaptation approaches for ABR are content-agnostic, without considering the semantic information of the video content.
Nevertheless, semantic information largely determines the informativeness and interestingness of the video content,
and consequently affects the QoE for video streaming.
One common case is that the user may expect higher quality for the parts of video content that are more interesting or informative so as to reduce video distortion and information loss, 
given that the overall bitrate budgets are limited.
This creates two main challenges for such a problem:
First, how to determine which parts of the video content are more interesting?
Second, how to allocate bitrate budgets for different parts of the video content with different significances?
To address these challenges, we propose a Content-of-Interest (CoI) based rate adaptation scheme for ABR.
We first design a deep learning approach for recognizing the interestingness of the video content,
and then design a Deep Q-Network (DQN) approach for rate adaptation by incorporating video interestingness information.
The experimental results show that our method can recognize video interestingness precisely,  
and the bitrate allocation for ABR can be aligned with the interestingness of video content
while not compromising the performances on objective QoE metrics.
\end{abstract}
\begin{IEEEkeywords}
video streaming, rate adaptation, video content analysis, deep reinforcement learning, QoE
\end{IEEEkeywords}

\section{Introduction}
Online video has become one of the most popular applications on the Internet,
and global Internet video traffic will grow threefold between 2016 and 2021 \cite{networking2016forecast}.
However, user viewing experience still needs improvements due to unstable network conditions and limited bandwidth capacities, especially for the users of mobile streaming services.
Moreover, the growing number of viewers and the wide adoption of High-Definition (HD) videos in streaming services make bandwidth requirements grow explosively.
This may further deteriorate user viewing experiences if the deployment of network resources cannot catch up with the growing demands of video consumption.
These realities make it challenging for video service providers to provide satisfactory viewing experiences.

Adaptive Bitrate (ABR) streaming is currently the most effective solution for video streaming under unstable network conditions.
Each video is encoded into many representations of different bitrates for ABR streaming.
The client can dynamically select the most suitable representation according to the current network conditions.
As such, the rate adaptation mechanism is vital to the performance of ABR streaming.
To design proper rate adaptation approaches for improving Quality of Experience (QoE) \cite{brunnstrom2013qualinet} for ABR streaming,
QoE metrics should be defined first so as to quantitatively evaluate the performance of rate adaptation.
The most commonly adopted QoE metrics in ABR streaming include rebuffering time, average bitrate, video quality variation, etc.
These are {\em objective} QoE metrics, as they are based upon measured performance parameters of the video delivery system.

The objective QoE metrics neglect the viewer's subjective feelings as they experience the video delivered to them \cite{brunnstrom2013qualinet}.
The user subjective engagement with the streamed video depends on what is happening in the video. Not all segments of the video draw the same attention from the user. For instance, for a user watching a soccer game, there is high engagement when the action is near the goal, but low attention when a player fetches the ball out of bounds. We denote by {\em interestingness} the level of (subjective) engagement that the video draws from the user.
Currently, video content is delivered in networks as binary data and the semantic-level information of video content is ignored by rate adaptation schemes.
However, the semantic information of video content plays an important role on the user's subjective viewing experiences, e.g.,
influencing user attention and interest.
Therefore, it is also necessary to consider the subjective QoE metrics for optimizing QoE.

The human visual attention system is selective \cite{kosslyn1996image}, and the more interesting parts of the video content  draw more user attention.
Allocating more bitrate budgets for the interesting parts of video content can achieve higher viewing experiences and reduce the information loss caused by video distortion.
However, due to the complexity of video content and the subtlety of the user's interest towards video content,
it is challenging to analyze video content from the user's perspective and incorporate the information for rate adaptation.
To address these problems, we first design a deep learning based approach for analyzing the interestingness of video content.
Then, we design a Deep Q-Network (DQN) based approach for rate adaptation by incorporating video interest information.
The method can learn the optimal rate adaptation policy by jointly considering buffer occupancy, bandwidth, and the interestingness of video content.
We evaluate the performance of our method using real-world datasets.
%

The rest of this paper is organized as follows.
Section \ref{sec:related-work} presents the related works on rate adaptation schemes.
Section \ref{sec:system-design} presents the system design and workflows.
Section \ref{sec:video-interest-analysis} presents the deep learning based approach for interestingness recognition.
Section \ref{sec:dqn-rate-adaptation} introduces the DQN based approach for rate adaptation while considering video interestingness information.
Section \ref{sec:performance-evaluation} presents the performance evaluation of our proposed method.
Section \ref{sec:conclusion} concludes this paper.

\section{Related Work} \label{sec:related-work}

Many existing works have studied the rate adaptation problem by considering different influence factors or using different mathematical models for maximizing QoE.

Huang \emph{et al.} \cite{huang2015buffer} designed a buffer-based approach by considering the current buffer occupancy.
Li \emph{et al.} \cite{li2014probe} designed a client-side rate adaptation algorithm by envisioning a general probe-and-adapt principle.
Yin \emph{et al.} \cite{Yin:2015:CAD:2785956.2787486} proposed a Model Predictive Control (MPC) approach by jointly considering buffer occupancy and bandwidth.
Bokani \emph{et al.} \cite{bokani2015optimizing} and Zhou \emph{et al.} \cite{zhou2016mdash} adopted Markov Decision Process (MDP) for rate adaptation.
Spiteri \emph{et al.} \cite{spiteri2016bola} adopted Lyapunov framework to design an online algorithm to minimize rebuffering and maximize QoE, without requiring bandwidth information.
Qin \emph{et al.} \cite{qincontrol} proposed a PID based method for rate adaptation,
and Mao \emph{et al.} \cite{mao2017neural} adopted deep reinforcement learning for rate adaptation.
In this line of works, they mainly considered the objective QoE metrics, aiming to improve the performances on
rebuffering time, average bitrate, and video quality variation.

Cavallaro \emph{et al.} \cite{cavallaro2005semantic} showed that the use of semantic video analysis prior to encoding for adaptive content delivery reduces bandwidth requirements.
Hu \emph{et al.} \cite{hu2017semantic} proposed a semantics-aware adaptation scheme for ABR streaming by semantic analysis for soccer video.
Fan \emph{et al.} \cite{fan2015segment} utilized various features collected from streaming services to determine if a video segment attracts viewers for optimizing live game streaming.
Dong \emph{et al.} \cite{dong2018personalized} designed a personalized emotion-aware video streaming system based on the user's emotional status.
In this line of works, they considered different subjective factors for optimizing video streaming services to improve QoE.

\section{System Design} \label{sec:system-design}
%
%
\begin{figure}
\begin{center}

\epsfig{file=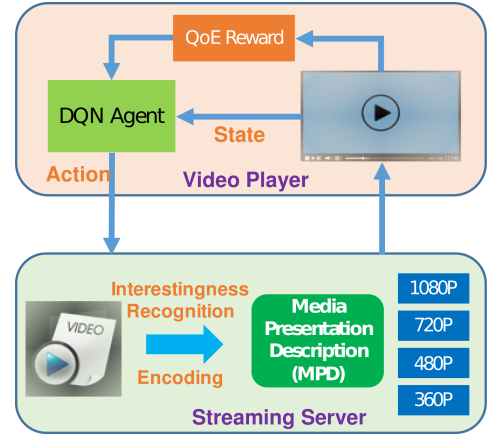, width=0.70\columnwidth}
\end{center}
\vspace{-1em}
\caption{The design of the COI based rate adaptation for ABR Streaming.}\label{fig:sys-design}
\end{figure}

We illustrate the design of the Content-of-Interest (CoI) based rate adaptation mechanism for ABR streaming in Fig. \ref{fig:sys-design}.
The system consists of the following components.

\emph{Streaming Server:}
The streaming server pre-processes video files and streams the video content to users.
For video pre-processing, each video file will be encoded into many representations at different bitrates and segmented into many equal-duration video chunks.
Each video chunk will be processed to analyze the interestingness of the video content.
The available bitrate information and the interestingness information of each video chunk will be included in the Media Presentation Description (MPD) manifest file \cite{stockhammer2011dynamic}.
In this work, we mainly consider Video-on-Demand (VoD) services, and the video encoding and interestingness recognition will be performed offline before video streaming.

\emph{Video Player:}
The video player requests the MPD of a video file when starting a video session and analyzes the available bitrates and the interestingness information of the video content.
The video player requests the selected video chunks from the streaming server, and measures the average bandwidth for downloading each video chunk.

\emph{DQN Agent:} We adopt the DQN method \cite{mnih2013playing} for rate adaptation.
The DQN agent will use the bandwidth, the current buffer occupancy, and the interestingness of the next several video chunks as the system state for determining which bitrate should be selected for the next video chunk.

\section{Interestingness Recognition Algorithm}\label{sec:video-interest-analysis}
In this section, we introduce the deep learning approach for recognizing the interestingness of video content.

\begin{figure}
\begin{center}{}
\epsfig{file=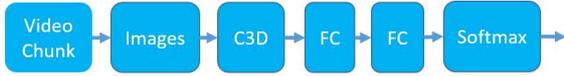, width=.85\columnwidth}
\end{center}
\vspace{-1em}
\caption{The deep learning model for video interestingness recognition.}\label{fig:video-interest-prediction}
\end{figure}

%
%
We illustrate the model for video interestingness recognition in Fig. \ref{fig:video-interest-prediction}.
Video chunks consi{}st of a series of video frames in time order.
It has been shown that 3D Convolutional Networks (3D ConvNets) are more suitable for learning spatiotemporal features \cite{tran2015learning},
there{}fore, we adopt 3D ConvNets for learning spatiotemporal features.
We extract 16 images from each video chunk and use 3D ConvNets to generate video features.
The extracted video features from 3D ConvNets will be input into two Fully-Connected (FC) layers, and the activation function for the fully-connected layers is Rectifier \cite{relu_function}.
The output layer has one node and the activation function is the Softmax function \cite{softmax_function}.
The output value is real-valued, which represents the interestingness of a video chunk,
and a higher value represents a higher level of video interestingness.
We adopt the TVSum dataset \cite{song2015tvsum} for training the network for interestingness recognition.
The dataset was created by segmenting videos into two second-long video segments,
and 20 users were invited to rate each segment compared to other segments from the same video.
The average of the rating for each segment is used as the ground truth,
and the scale is from one to five. 
The  data is split into small batches that are used to calculate the loss and update the network in each training epoch.
The loss function is the Mean Squared Error (MSE),
\begin{equation}\label{eqn:mse-loss}
MSE = \frac{1}{n} \sum_{i=1}^n (y_i - \hat{y_i})^2,
\end{equation}
where $n$ is the number of samples (video chunks) in each training batch, $\hat{y_i}$ is the predicted interestingness of sample $i$, 
and $y_i$ is the ground-truth of the interestingness of sample $i$.
For training the network, 
we adopt Adam \cite{kingma2014adam} for training the fully connected layers and the output layer.

%
%

\begin{table}
\centering
\caption{Key Notations and Definitions} \label{tabel:key-notation}
\begin{tabular}{p{1.45cm}|p{6.2cm}} \hline
Notation                & Definition       \\ \hline
$t$                     & the discrete time slot, $t = 1,2, ...$         \\
$s_t, a_t, r_t$         & system state, action, reward at time slot $t$ \\
$B$                     & the set of available bitrates for each video \\
$v_t$                   & the average bandwidth for downloading video chunk $t$\\
$I_t$                   & the interestingness of video chunk $t$ \\
$\overrightarrow{v_t}$  & the vector of the average bandwidth for downloading the next $k$ video chunks\\
$L_t$                   & buffer occupancy before downloading video chunk $t$ \\
$b_t$                   & the selected bitrate for video chunk $t$ \\
$\overrightarrow{w_t}$  & the vector consisting of the interestingness of the following $h$ video chunks \\
$\pi$                   & the policy for choosing bitrate for the next video chunk \\
$r_t$                   & reward during time slot $t$ \\
$f(\cdot)$              & mapping the interestingness of a video chunk to the weight for a video chunk \\
$q(\cdot)$              & mapping video bitrate to video quality \\
$\alpha$                & the weight for the penalty of rebuffering time \\
$\beta$                 & the weight for the penalty of quality variation \\
$Q(s,a)$                & the quality of the state-action combination\\
$N$                     & the number of transitions chosen from replay buffer for minibatch training\\
$\theta$                & the weights of the DQN network\\
\hline\end{tabular}
\end{table}

\section{DQN based Interest-Aware Rate Adaptation} \label{sec:dqn-rate-adaptation}
In this section, we introduce the DQN based interest-aware rate adaptation for ABR streaming.
The key notations used in this paper are summarized in Table \ref{tabel:key-notation}.
%

\subsection{Problem Formulation for Interest-Aware Rate Adaptation}
We adopt a discrete time system, where the time is denoted as $t=1,2,3,...$.
The duration of each time slot may not be equal, and depends on the time for downloading a video chunk.
We formulate the interest-aware rate adaptation as a Reinforcement Learning (RL) problem,
where the agent interacts with the streaming environment for learning the optimal rate adaptation policy.
More specifically, after downloading video chunk $t-1$, the agent receives the observed system state $s_t$,
then takes action $a_t$ for selecting the bitrate for video chunk $t$ according to the current policy, and finally gets reward $r_t$ after downloading video chunk $t$.
These procedures will be repeated until the end of a video session.

\textbf{Streaming Environment:}
We denote the set of available bitrates in the streaming system for each video as $B$.
The bandwidth during a video session is time-varying, and we denote the average bandwidth for downloading video chunk $t$ as $v_t$.
The interestingness of video chunk $t$ is denoted as $w_t$.
The selected bitrate for video chunk $t$ is denoted as $b_t$.
%
%

\textbf{State:}
The state describes the bandwidth of the streaming service, the buffer occupancy of the video player, and the interestingness of the following video chunks, etc.
We denote the state at time slot $t$ as $s_t$, specifically,
\begin{equation}\label{eqn:state}
s_t = (\overrightarrow{v_t}, L_t, b_{t-1}, \overrightarrow{w_t}, \overrightarrow{u_t}),
\end{equation}
where $\overrightarrow{v_t}$ is the vector consisting of the predicted average bandwidth for downloading the next $k$ video chunks (i.e., $\overrightarrow{v_t} = (v_{t},v_{t+1}, ...,v_{t+k-1})$),
$L_t$ is the buffer occupancy before downloading video chunk $t$,
$b_{t-1}$ is the selected bitrate for video chunk $t-1$,
$\overrightarrow{w_t}$ is the vector consisting of the interestingness of the following $h$ video chunks (i.e., $\overrightarrow{w_t} = (w_{t},w_{t+1}, ...,w_{t+h-1})$),
$\overrightarrow{u_t}$ is the vector consisting of the available chunk sizes of video chunk $t$.
Here, the interestingness information for each video chunk of a whole video file is known at the start of a video session,
because video content will be pre-processed on the server and the interestingness information will be included in MPD.

\textbf{Action:}
The control action for the agent is to select the bitrate for the next requested video chunk according to the current system state, which can be described as
\begin{equation}
a_t = \pi(s_t) \to b_t, b_t \in B,
\end{equation}
where $\pi$ is the policy for selecting bitrate.

\textbf{Reward:}
%
%
%
We adopt the following utility function revised based on the QoE metrics defined in \cite{Yin:2015:CAD:2785956.2787486} for measuring the reward during a time slot,
\begin{equation} \label{eqn:qoe-function}
r_t(s_t, a_t) = \underbrace{f(w_t)}_{\text{weight}} \underbrace{q(b_t)}_{\text{quality}} - \underbrace{\alpha R_t}_{\text{video stall}} - \underbrace{\beta |q(b_t) - q(b_{t-1})|}_{\text{quality variation}},
\end{equation}
where $r_t$ is the reward for time slot $t$,
$f(\cdot)$ maps the interestingness of a video chunk to the weight for a video chunk,
$q(\cdot)$ maps video bitrate to video quality,
$\alpha$ is the weight for the penalty of rebuffering time,
$R_t$ is the rebuffering time incurred during time slot $t$,
and $\beta$ is the weight for the penalty of quality variations.
With the reward function in Eq. \ref{eqn:qoe-function}, the video chunks with higher interestingness have higher weights,
therefore, the agent will get more rewards if the video chunks with higher interestingness are allocated more bitrate budgets.

\textbf{Objective:}
Our objective is to derive the optimal rate adaptation policy for maximizing the rewards over a video session. Due to the uncertainly of system dynamics, future rewards and present rewards have different importance and weights.
Therefore, we maximize the overall discounted rewards,
in which the present rewards have higher importance and the future rewards have less importance, mathematically,
\begin{equation}\label{eqn:qoe-maximization}
\pi^{*} = \argmax_{\pi} \mathop{\mathbb{E_\pi}} \sum_{i=0}^{\infty} \gamma^{i} r_{t+i}(s_{t+i}, a_{t+i}),
\end{equation}
where $\pi^{*}$ is the optimal rate adaptation policy that needs to be derived and $\gamma$ is the discount factor.

\subsection{DQN for Learning Rate Adaptation Policy}
\begin{figure}
\begin{center}
\epsfig{file=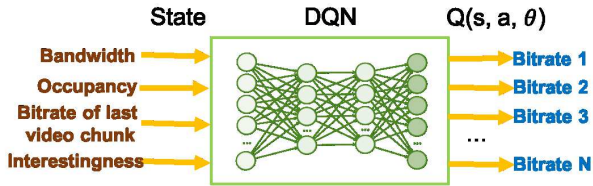, width=.9\columnwidth}
\end{center}
\vspace{-1em}
\caption{The DQN network for interest-aware rate adaptation.}\label{fig:dqn-based}
\end{figure}
We adopt DQN \cite{mnih2013playing} for learning the rate adaptation policy, and the network of DQN is illustrated in Fig. \ref{fig:dqn-based}.
The inputs of the network are the system states listed in Eq. \eqref{eqn:state},
and the outputs of the network are the action-value function, $Q(s, a, \theta)$,
which represents the quality of the state-action combinations for each state $s$ and action $a$. 
$\theta$ represents the weights of Q network, which will be updated during training.

We illustrate the details of the DQN based learning algorithm for rate adaptation in Algorithm \ref{DQN_solution}.
At the start of each video session, the video player is initialized and a video file is randomly chosen.
When selecting the bitrate for a video chunk, the agent randomly selects a bitrate with probability $\epsilon$.
Otherwise, the agent will choose the bitrate that has the maximum action-value given the current state.
The video player will download the video chunk of the selected bitrate.
After the completion of the download, the agent will calculate the reward according to Eq. \eqref{eqn:qoe-function}
and observe the next state.
The transition $(s_t, a_t, r_t, s_{t+1})$  will be stored into the replay buffer.
We will randomly choose N transitions from replay buffer for training the network at each gradient descent step.
For each sampled transition, we denote it as $(s_{t'}, a_{t'}, r_{t'}, s_{t'+1})$.
The following loss function is adopted for training DQN,
\begin{equation} \label{eqn:loss-function}
L(\theta_i) =  \mathbb{E}[(y_{t'} - Q(s_{t'},a_{t'};\theta_i))^2],
\end{equation}
where $y_{t'} = \mathbb{E}[r_{t'}+\gamma \max_{a'}Q(s_{t'+1},a';\theta_{i-1})|s_{t'},a_{t'}]$ and
$\theta_i$ denotes the weights of the Q network at the $i$-th iteration.
Then, a mini-batch gradient descent step will be performed to update the weights of the Q network.

After the training, the Q network will be adopted by the agent for making rate adaption decision.
For the next requested video chunk, the bitrate which has the largest action-value for the current state will be selected by the agent.
\begin{algorithm}
\renewcommand{\algorithmicrequire}{\textbf{Input:}}
\renewcommand\algorithmicensure {\textbf{Output:} }
\caption{DQN for Interest-Aware Rate Adaptation} \label{DQN_solution}
\begin{algorithmic}[1]
\State{Initialize replay memory D}
\State{Initialize Q Network with random weights}
\For{video session $ = 1,2,...,M$}
\State{Initialize the video player and choose a video file}
\State{Observe initial state $s_1$}
\For{video chunk $t = 1,2,...,K$}
\State{With probability $\epsilon$ randomly select a bitrate $a_t$}
\State{otherwise select bitrate $a_t = \argmax_a Q(s_t, a; \theta)$}
\State{Download video chunk $t$ until completed}
\State{Observe reward $r_t$ and next sate $s_{t+1}$}
\State{Store transition $(s_t, a_t, r_t, s_{t+1})$ into $D$}
\State{Randomly sample $N$ transitions from D}
\State{Set $y_{t'} = r_{t'}$, if the video session ends}
\State{otherwise set $y_{t'} = r_{t'} + \gamma \max_{a} Q(s_{t'+1},a;\theta_{i-1})$}
\State{Train the network using Eq. \eqref{eqn:loss-function} as loss function}
\EndFor
\EndFor
\end{algorithmic}
\end{algorithm}
\section{Experiment} \label{sec:performance-evaluation}
In this section, we illustrate the experiment settings and the performance of the CoI based rate adaptation method.
\subsection{Experimental Settings}
To simulate different network conditions, we adopt the FCC broadband dataset \cite{fcc_dataset}
and the 3G/HSDPA mobile dataset \cite{riiser2013commute} for training DQN and evaluating performance.
In our experiment, $\overrightarrow{v_t}$ is the vector of the predicted bandwidth for the next two video chunks. 
$\overrightarrow{w_t}$ is the vector of the video interestingness for the next three video chunks.
We adopt the settings of the penalty for rebuffering time and quality variations used in  \cite{Yin:2015:CAD:2785956.2787486},
where $\alpha$ is 3000, $\beta$ is 1, and $q(\cdot)$ are identity functions.
$f(\cdot)$ scales the video interestingness values from 1-5 to 1-3 with normalization.
The available bitrate levels are 350kbps, 600kbps, 1000kbps, 2000kbps, 3000kbps.
%

For the DQN agent, after the hyper-parameters searching and tuning, we adopt the following parameters setting: we use a fully-connected neural network with two hidden layers of size 256 and 512, the activation function is ReLu, and the output layer uses a linear activation function to output the approximated Q value for a given state and action pair. A naive $\epsilon$-greedy policy is used for exploration and the probability of randomly selecting an action during training is 0.2. The learning rate is 0.1, the replay buffer size of DQN is 10000, the discount factor is 0.8, the decay parameter for updating target Q network is 0.5, the batch size is 256, and for each instance of training, we sample 50 batches of data.

\subsection{Baseline Methods}
We compare the performances of our method with the following methods:
1) Buffer-Based (BB) approach \cite{huang2015buffer} chooses the bitrate for the next video chunk as a function of the buffer occupancy.
In our settings, the reservoir (r) is five seconds and the cushion (c) is 20 seconds.
2) Rate-Based (RB) approach chooses the maximum available bitrate less than the predicted bandwidth.
3) Robust-MPC approach \cite{Yin:2015:CAD:2785956.2787486} uses MPC method to select the bitrate for maximizing the overall QoE over the prediction horizon.
The prediction horizon of Robust-MPC is three time slots.
4) DQN-Constant approach also adopts DQN method for rate adaptation, however, the weights of the video chunk is constantly set as two. 
RB, Robust-MPC, DQN-Constant, and our proposed approach use the harmonic mean of the average bandwidth of the past 5 video chunks as bandwidth prediction for the next video chunk.

\subsection{Performance Evaluation}
\subsubsection{Video Interestingness Recognition Precision}
\begin{figure}
\begin{center}
\epsfig{file=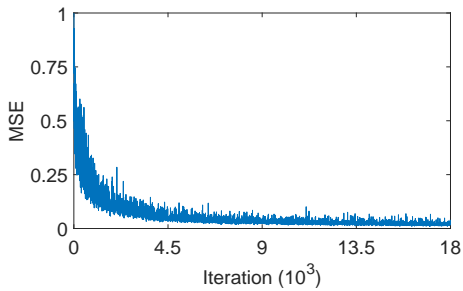, width=.7\columnwidth}
\end{center}
\vspace{-1em}
\caption{The interestingness recognition error during different iterations. The recognition error is converged to 0.02 after 18,000 iterations. }\label{fig:mse-prediction}
\end{figure}
\begin{figure}
\begin{center}
\epsfig{file=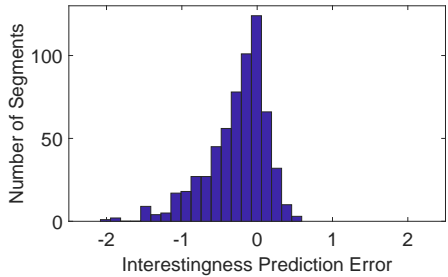, width=.7\columnwidth}
\end{center}
\vspace{-1em}
\caption{The interestingness recognition error distribution. The error is mainly distributed around 0.0 which demonstrates the good performance of the recognition model.}\label{fig:error-dis}
\end{figure}
\begin{figure}
\begin{center}
\epsfig{file=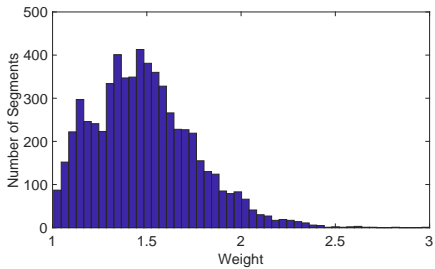, width=.7\columnwidth}
\end{center}
\vspace{-1em}
\caption{The distribution of the weights of the video chunks. After scaling the interestingness values from [1.0, 5.0] to [1.0, 3.0], the weights are mainly distributed from 1.0 to 2.0 with the mean at around 1.5.}\label{fig:interest-distribution}
\end{figure}
There are overall 6245 user-annotated video chunks in the dataset, and we randomly choose 90\% of the video chunks for training and 10\% of the video chunks for evaluating the performance.
In Fig. \ref{fig:mse-prediction}, we illustrate the interestingness recognition error during different iterations in the training stage.
It can be observed that the recognition error decreases over the training iterations and finally converges,
and the MSE converges to 0.02 after 18,000 iterations.
The interestingness recognition error distribution is illustrated in Fig. \ref{fig:error-dis}, and the mean error is 0.34.
The interestingness prediction is biased towards giving a lower score, because the interestingness values of most of the video chunks are small, 
and the prediction algorithm tends to predict a lower value for reducing the overall MSE.
We use the normalization function as $f(\cdot)$ in Eq. \eqref{eqn:qoe-function} for scaling the interestingness value into the weight of a video chunk.
The range of the weight is from 1.0 to 3.0.
The overall distribution of the weights of the video chunks is illustrated in Fig. \ref{fig:interest-distribution}.

\subsubsection{Performances on Rebuffering Time, Average Bitrate, and Bitrate Variations}
We first evaluate the performance of different methods on rebuffering time, bitrate variation, and video quality.
We run the tests over 40 video sessions, and each video session has 200 video chunks.
For each video session, we randomly choose a bandwidth trace and the interestingness information of a video file.
The performance of each method is illustrated in Table \ref{tab:table2}.
From the results in Table \ref{tab:table2}, we can observe that the performances of our proposed CoI method on rebuffering time,
average bitrate, and quality variations are close to the performances of the state-of-the-art methods, including Robust-MPC, BBA, and RBA. 
This verifies that introducing video interestingness information for rate adaptation will not deteriorate the performances from the perspective of objective QoE metrics.
Moreover, CoI reaches the highest mean value of average bitrate per session out of all the methods and the lowest standard deviation of it.
For average rebuffering time, the CoI method is lower than the BBA and close to the Robust-MPC.
For the bitrate variation, CoI method is lower than the BBA and quite close to the Robust-MPC.

Note that the average bitrate and rebuffering time will both increase under the CoI method. 
This is due to that the video interestingness value is larger than one, and it will increase the weight of video quality in the reward function (Eq. \eqref{eqn:qoe-function}),
compared with rebuffering time and quality variations.
For verification, we can observe that DQN-Constant has a higher average bitrate compared with Robust-MPC, BBA, and RBA, yet the rebuffering time of DQN-Constant is also significantly larger than the other methods.

\begin{figure}
\begin{center}
\epsfig{file=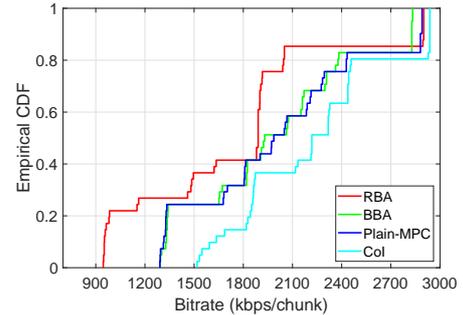, width=.7\columnwidth}
\end{center}
\vspace{-1em}
\caption{The Empirical CDF of average bitrate per session. The results show that CoI tends to allocate a higher bitrate for each video chunck.}\label{fig:cdf-bitrate}
\end{figure}

\begin{figure}
\begin{center}
\epsfig{file=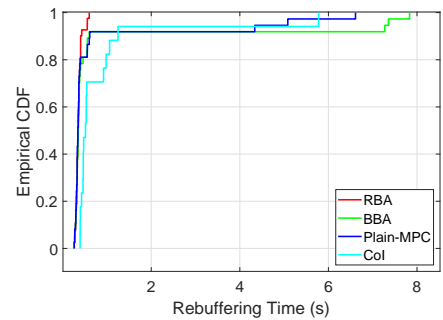, width=.7\columnwidth}
\end{center}
\vspace{-1em}
\caption{The Empirical CDF of average rebuffering time per session. It can be observed that CoI method maintains a relatively low rebuffering time even under higher bitrate selection comparing to other methods.}\label{fig:cdf-rebuffering}
\end{figure}

\begin{figure}
\begin{center}
\epsfig{file=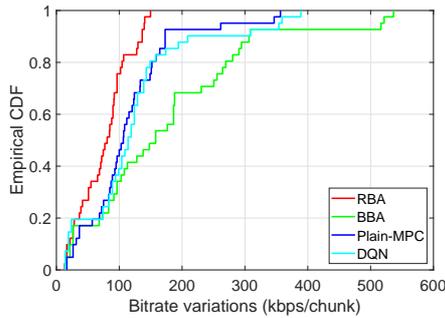, width=.7\columnwidth}
\end{center}
\vspace{-1em}
\caption{The Empirical CDF of average bitrate variations per session. Though quality variation only accounts for a small part of the reward, the CoI method still controls the bitrate variation to a level comparable to the other methods.}\label{fig:cdf-switching}
\end{figure}

We also give the empirical distributions of average bitrate, rebuffering time, and quality variations of different methods in Fig. \ref{fig:cdf-bitrate}, \ref{fig:cdf-rebuffering}, and \ref{fig:cdf-switching}. We can observe that the CoI method has the highest distributions on bitrate comparing with the rest methods. For the distributions of rebuffering time and quality variations, the CoI method gets quite good results though not the lowest since there is a trade-off between minimizing the rebuffering time, quality variations and maximizing the video interestingness value.

\begin{table*}[t]
  \centering
  \caption{The average performances per video session.}
  \label{tab:table2}
  \begin{tabularx}{\textwidth}{  p{6cm}  p{1.8cm}  p{1.8cm}  p{1.8cm}  p{1.8cm} p{1.8cm}}
  \toprule

                                                        & RBA       & BBA      & Robust-MPC & CoI     & DQN-Constant\\
    \midrule
  \rowcolor{LightCyan}
  Average Rebuffering Time (s)                          & 0.3617    & 0.9439   & 0.7661     & 0.9173  & 1.915      \\
  Standard Deviation of Rebuffering Time (s)            & 0.0717    & 1.9731   & 1.4079     & 1.2803  & 2.397      \\

  \rowcolor{LightCyan}
  Average Bitrate(kbps)                                 & 1762.3    & 1996.6   & 2014.5     & 2231.8  & 2145.6     \\
  Standard Deviation of Average Bitrate(kbps)           & 617.1     & 517.7    & 538.5      & 452.4   & 512.9      \\

  \rowcolor{LightCyan}
  Bitrate Variation (kbps/chunk)                        & 76.3598    & 176.5488   & 115.5366     & 124.5122  & 202.183       \\
  Standard Deviation of  Bitrate Variation (kbps/chunk) & 39.5099    & 133.6111   & 74.3199     & 91.2549    & 162.813      \\

    \bottomrule
  \end{tabularx}
\end{table*}

\subsubsection{Relation between Video Interestingness and Average Bitrate}
\begin{figure}
\begin{center}
\epsfig{file=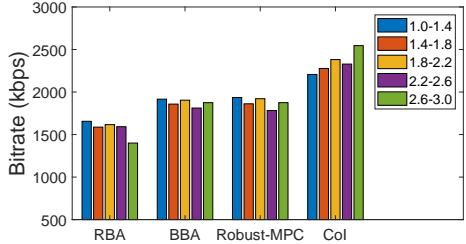, width=.7\columnwidth}
\end{center}
\vspace{-1em}
\caption{The average bitrates for different levels of video interestingness. It can be observed that CoI method tends to allocate more birtrate budgets to video chunks that have higher video interestingness whereas other methods don't show the tendency.}\label{fig:Interest-level}
\end{figure}

\begin{figure}
\begin{center}
\epsfig{file=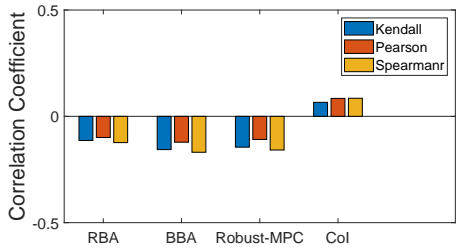, width=.7\columnwidth}
\end{center}
\vspace{-1em}
\caption{The correlation coefficient between video interestingness and  bitrate. The results show that there is no linear correlation between video interestingness and bitrate for RBA, BBA, and Robust-MPC methods. But the result of CoI method shows a positive correlation.}\label{fig:Interest-correlation}
\end{figure}

We illustrate the average bitrate for different levels of video interestingness in Fig. \ref{fig:Interest-level}.
Because video interestingness is real-valued, we divide the interestingness of the video chunks into four levels, namely,
1.0-1.4, 1.4-1.8, 1.8-2.2, 2.2-2.6 and 2.6-3.0.
We can observe that the average bitrates for the video chunks with higher levels of interestingness are allocated with higher bitrate budgets on average.
This verifies the effectiveness of the DQN method for aligning bitrate allocation with video interestingness.
In comparison, the other content-agnostic rate adaptation methods, which ignore video interestingness information, will allocate the bitrate budgets equally among different levels of video interestingness.
We also evaluate the correlation between video interestingness and average bitrate for different methods using Pearson coefficient, Spearman coefficient, Kendall's tau coefficient, and the results are shown in Fig. \ref{fig:Interest-correlation}.
The results show that there is no linear correlation between the variables for the content-agnostic approaches.
In contrast, the average bitrate and video interestingness are positively correlated with each other under the CoI method.

\subsubsection{Convergence of DQN agent with different hyper-parameters setting}
\begin{figure}
\begin{center}
\epsfig{file=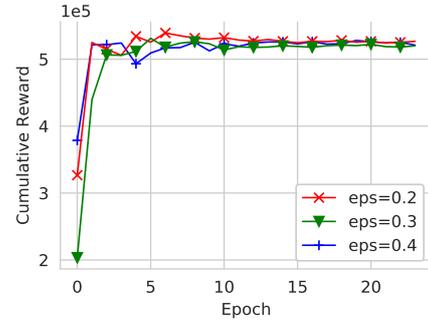, width=.7\columnwidth}
\end{center}
\vspace{-1em}
\caption{The average cumulative rewards of DQN agent under different probability of $\epsilon$-greedy strategy. The DQN agent gets the highest cumulative rewards with the probability of 0.2 to randomly choose the actions.}\label{fig:Q-reward}
\end{figure}

We also verify the convergence of DQN agent with different hyper-parameters setting, including the network size, learning rate, exploration strategy etc. All the results prove the robustness of our DQN agent with the environment. 
Fig. \ref{fig:Q-reward} shows the cumulative reward of the DQN agent with different $\epsilon$-greedy strategies. It can achieve the best performance when $\epsilon$ is 0.2.

\section{Conclusion} \label{sec:conclusion}
In this work, 
%
%
we proposed a CoI based rate adaptation method for ABR streaming.
We first developed a deep learning method for recognizing the interestingness of the video content,
and then developed a DQN method which can incorporate interestingness information for rate adaptation so that
the video content with higher interestingness will be allocated with higher bitrate budgets.
Compared with the state-of-the-art rate adaptation methods, 
the CoI method will not compromise the performances on the objective QoE metrics of average bitrate, rebuffering time, and quality variations.
Therefore, it can have more advantages compared with the content-agnostic rate adaptation methods in some video streaming scenarios.

Our method has the following limitations.
First, different application scenarios may have different criteria for video interestingness. 
For instance, in video lectures, the informativeness of the video content may determine its interestingness to the viewers;
in sport videos, the interestingness may be determined by the actions being played.
Second, users may require different video quality differentiation among the video content of different levels of interestingness.
For instance, in some scenarios, the user may only require a slightly higher quality for the video content with higher interestingness, 
while in other scenarios the user may require a significant higher quality. 
These problems require the CoI method to be customized according to the specific requirements of a given scenario, e.g., implementing dataset for training the interestingness prediction algorithm or tuning the DQN model to achieve the required quality differentiation.
Nevertheless, our method has the elasticity for achieving the personalization.

\bibliographystyle{IEEEtran}
\bibliography{ICC-2019-Mozart_arxiv}

\end{document}